# Transparent conductive oxides and low loss nitride-rich silicon waveguides as building blocks for neuromorphic photonics

J. Gosciniak[1*] and Jacob B. Khurgin[2]

[1]*ENSEMBLE3 sp. z o.o., Wolczynska 133, 01-919 Warsaw, Poland*
[2]*Electrical and Computer Engineering Department, Johns Hopkins University, Baltimore, Maryland 21218, USA*

Corresponding authors: * jeckug10@yahoo.com.sg

**Abstract**
Fully CMOS-compatible photonic memory holding devices hold a potential in a development of ultrafast artificial neural networks. Leveraging the benefits of photonics such as high-bandwidth, low latencies, low-energy interconnect and high speed they can overcome the existing limits of the electronic processing. To satisfy all these requirements a new photonic platform is proposed that combines low-loss nitride-rich silicon as a guide and low-loss transparent conductive oxides as an active material that can provide high nonlinearity and bistability under both electrical and optical signals.

**Introduction**
Neuromorphic computing refers to the way the signal is processed that try to mimic a signal processing by a brain [**1**]. In comparison to traditional computers that are based on von Neumann architecture with two separated memory and processing units and operating in a sequential way [**2**], the brain process signals in a parallel way [**3, 4**]. It provides huge benefits in terms of speed and energy efficiency as a data transfer is responsible for a large part of the power consumption. One of the ways to overcome some of those limitations is by developing new algorithms that can improve signal processing [**5, 6**], however, it still requires the data transfer between memory and processor what limits its efficiency. To deal with those limitations a lot of effort is put in the last years in a development of artificial neurons and synapses that can be implemented in the network [**1**].

Neuromorphic computing based on the photonics *i.e.*, the neuromorphic photonics, avail photons as a signal carrier to transfer an information between different part of the network [**7-12**]. Thanks to almost unlimited bandwidth, compatibility with standard CMOS technology and almost zero power consumption to carry out basic matrix multiplication it can offer a huge improvement compared to neuromorphic electronics. The full parallelism can be achieved by busing multiple signals on a single waveguide at the speed of light. Simultaneously, the optical weights can offer low latency of the computation. By combining those advantages at least few orders of magnitude improvement compared to electronic counterparts is expected. However, a realization of such demanding task requires new material platform and low-loss architecture that is still missing.

Silicon nitride (SiN) is a ubiquitous material for photonic integrated circuit (PIC) technologies since it is compatible with standard CMOS processes [**13, 14**]. It allows for cost-effective construction of devices and co-integration of electronic and photonics components on a single chip. Furthermore, the photonic devices based on SiN platform are characterized by higher tolerance to the temperature drifts compared to other materials, lower optical losses and broad wavelength range operation, wider wavelength transparency and improved crosstalk values [**14**]. Already, SiN has proved to be a proper material platform for a realization of neural networks showing the increased degree of freedom is design linear neurons [**8, 9**]. Thus, SiN platform can play a key role as routing layer in neuromorphic photonics [**9**].

Among many active materials available for implementation in neuromorphic networks [1], transparent conductive oxides (TCOs) seem to be a material of choice for such tasks as it provides nonlinearity and bistability under both electrical signal and optical power coupled to the waveguide [15, 16]. Thus, it can provide a dual-mode operation and bring a lot of flexibility in terms of operation conditions [17]. As it has already shown [15, 16], it exhibits two stable states that depends on the history of the system, thus it can act as a memristor [18-21]. TCOs belong to the epsilon-near-zero (ENZ) materials that show large permittivity tunability under an applied voltage and/or light illumination [17, 22-24]. They are characterized by fast switching time and low switching voltage when operating under an electrical switching mechanism [17, 22-24] what is a huge benefit for a realization of efficient neuromorphic networks. And similarly to the SiN platform, TCOs are CMOS-compatible and operate under low optical loss. Thus, a combination of SiN and TCOs can provide an ideal material platform for a realization of low-loss, CMOS-compatible and extremally fast neuromorphic systems able to process an information in-place and under a low operation power.

To process all information in-place, system has to possess some type of bistability, *i.e.*, a special activity that take place in biological neurons where neurons can switch between active and non-active states under some action of neuromodulating substances [3]. Thus, bistability is a property of a system that exhibits two stable steady states and the system rests in one of those states depending on the history of this system [15, 16, 18, 25-32]. It can refer to two opposite magnetizations of magnet, low or high resistance of the electronic devices, low or high signal transmitted through a device and etc. The two states represent two values of a binary digit *i.e.*, bit. To meet the demands of modern systems they should operate at high speed, under low power consumption and in wide operation bandwidth. However, up to now, most of the proposed bistable devices suffer either from high power consumption, incompatibility with standard processing technology, narrow bandwidth or complicated design that is a combination of a nonlinear material and resonant cavity [31]. The reliable bistable all-optical devices can bring progress in many fields, especially, in the all-optical neural networks, thus, the search for such device intensify in the last few years [15, 16, 31, 32].

**Switching mechanism**

The photonic devices with TCO materials can operate in dual-mode operation, electrical and/or optical, thanks to the unique properties of TCO materials which exhibit a dispersion of its real electrical permittivity under applying an electric field or optical pump, thus, either generating or exciting free carriers [17, 22]. Depending on the requirements and working conditions each of the processes can be implemented to the proposed device.

Electrical switching

Under an applied voltage the electrons accumulate at the TCO what increases the local density of electrons and reduces the permittivity according to the Drude dispersion formula:

$$\varepsilon(\omega) = \varepsilon_\infty - \frac{N_c e^2}{\varepsilon_0 (\omega^2 + i\omega\gamma) m^*(E)}$$

where $\varepsilon_\infty$ is the permittivity due to the bound electrons, $N_c$ is the carrier density, $e$ is the electron charge, $\omega$ is the working frequency, $m^*(E)$ is the energy-dependent effective mass, and $\gamma$ is the scattering rate. As it has been previously showed [33], even unity order permittivity change can be obtained under a reasonable voltage. The increased carrier concentration leads to decreasing a permittivity and shifting into the ENZ region what leads to higher absorption and increases the absorption losses of a device as the mode is more confined to the TCO material. Once a voltage is removed, electrons flow away from the TCO and TCO returns to its initial low-loss state. It should be emphasized that the switching process under the electrical modulation is limited by RC delays that scale with device size [34].

All-optical switching

In comparison, all-optical switching with TCO operates via two mechanisms, either via interband absorption, or through intraband absorption of light. For interband absorption, the energy of the optical pump has to be greater than the bandgap of the TCO to excite photocarriers from the valence band to the conduction band [**17, 23**]. As in a case of the electrical switching, the photoexcited carriers lower the permittivity of the TCO via Drude dispersion and move TCO closer to the ENZ region. On the other hand, intraband absorption with the pump energy lower than the bandgap, heats up electrons in the conduction band what move it toward higher energies. Due to the non-parabolic nature of the conduction bands in TCO, these excited electrons have a greater effective mass. For a Drude formula it can be seen that as the effective mass of electrons increase the plasma frequency decreases and, in consequence, it leads to increases of the TCO permittivity. When the optical pump is off, the electrons cool down in sub-picosecond time scale. Thus, all-optical switching is a very promising mechanism for realization of active photonic components operating in the femtosecond time scale. Furthermore, when operating under intraband absorption, the same light source can be used as a pump source and signal what reduces the complexity of the system.

**Design**

Here we examine a concept of bistability in a SiN rib photonic waveguide arrangement with TCO placed between SiN rib and ridge and utilizing an intraband absorption of light. Compared to our previous papers [**15, 16**] in which we utilized plasmonic slot waveguides to enhance an electric field into TCO and thus enhances the interaction of light with TCO, here we focused on all-dielectric device. It may provide lower electric field enhancement inside TCO but simultaneously it facilitates integration with photonic platform as it will not require any additional fabrication steps. Furthermore, a coupling efficiency between photonic waveguide and plasmonic slot waveguide usually does not exceed 50 %. In comparison, a proposed all-dielectric device can be easily integrated with the SiN photonic platform with an extremally high coupling efficiency exceeding 95% and it not require any additional fabrication steps

Here, the concept of bistability was investigated using 2D finite element method (FEM) simulations at the telecom wavelength of 1550 nm using a commercial software COMSOL and Lumerical. The thickness of TCO was chosen at 10 nm, while the thickness of SiN rib at 200 nm. The thickness and width of the SiN ridge was taken at $h$=300 nm and $w$=500 nm. The refractive index of SiN is assumed to be $n$=1.9963. For all TCOs considered here, the calculations were performed for a thermalization time $\tau$ = 500 fs. The ITO properties were taken as $\omega_p$=2.52·10$^{15}$ (rad/s), $\gamma$=1.8·10$^{14}$ (rad/s), $\varepsilon_\infty$=3.9 [**11**] where $\omega_p$ is the plasma frequency, $\gamma$ is the scattering rate and $\varepsilon_\infty$ is the permittivity due to the bound electrons. Similarly, the 6% Ga:ZnO (GZO) properties were taken at $\omega_p$=2.93·10$^{15}$ (rad/s), $\gamma$=1.78·10$^{14}$ (rad/s), $\varepsilon_\infty$=2.475 [**35**], 10 % Al:ZnO (AZO) at $\omega_p$=1.137·10$^{15}$ (rad/s), $\gamma$=1.27·10$^{14}$ (rad/s), $\varepsilon_\infty$=3.8825 [**36**] while In doped CdO at $\omega_p$=2.41·10$^{15}$ (rad/s), $\gamma$=3.06·10$^{13}$ (rad/s), $\varepsilon_\infty$=5.5 [**37-39**].

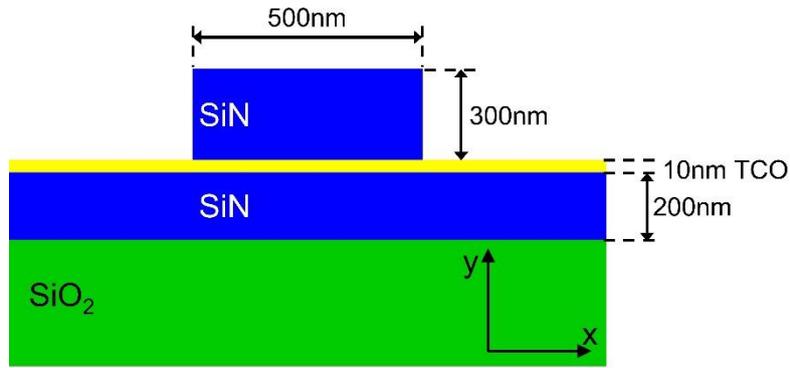

**Figure 1.** Geometry of the proposed photonic bistable device.

The TCO parameters presented above allowed to calculate wavelength and plasma frequency dependent complex permittivity of all TCO materials examined in this paper.

In our previous papers we focused on ITO [**15, 16**] as it is currently the most popular TCO material commonly found in a literature [**17, 22-24, 32-34**]. However, the family of TCO materials is very broad and, depending on applications and an operation wavelength range, the proper TCO material can be identify. In this paper we examine first four TCO materials: AZO, GZO, ITO and In doped CdO that represents wide spectrum of ENZ wavelengths ranging from *λ*=1.0 μm for 6 % Ga:ZnO (GZO) through, *λ*=1.5 μm for ITO, *λ*=1.82 μm for In:CdO to *λ*=3.34 μm for 10 % Al:ZnO (AZO). As observed, AZO and In:CdO are characterized by the lowest imaginary permittivity, thus losses, while the imaginary part of permittivity of ITO at ENZ wavelength is pretty high (**Fig. 2a**). As we are here interested in telecom wavelengths, in the rest of the paper we focus on GZO, ITO and In:CdO (**Fig. 2b**). GZO shows the lowest plasma frequency in the telecom wavelength of 1550 nm while the plasma frequency of In:CdO is the highest. However, as in previous case (**Fig. 2a**), the imaginary part of permittivity is lowest for In:CdO (**Fig. 2b**). Furthermore, it should be remembered that In:CdO is characterized by an order of magnitude higher mobility compared to any other TCOs what highly influences its scattering rate, damping factor, *γ* (*γ*=e/*μ*$m_{eff}$) where *μ* is the material mobility [**37, 38**].

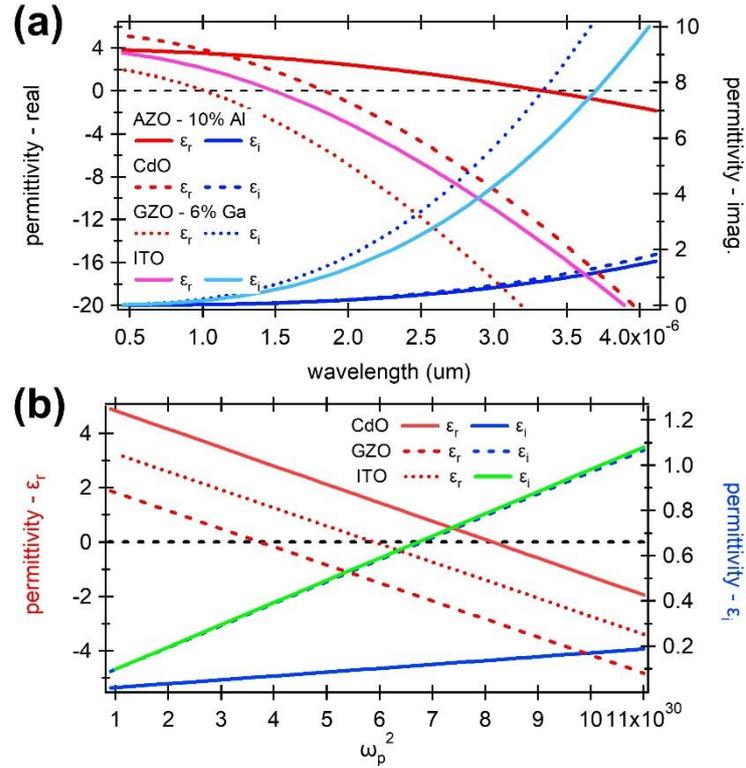

**Figure 2.** (a, b) Dispersion of real and imaginary parts of dielectric permittivity of ITO, GZO, AZO and In:CdO as a function of wavelength and plasma frequency.

By comparing a real part of permittivity in a function of wavelength it can be observed that change of real part of permittivity close to ENZ wavelength is higher for Indium doped CdO and GZO compared to ITO and AZO. Similarly, AZO is characterized by the smoothest transition close to the ENZ wavelength. From our previous papers we can deduce [15, 16] that the steeper slope of the permittivity close to the ENZ point the narrower absorption curve of the device what means that less power is required to switch between two transmission levels of a bistable device.

As observed from **Fig. 3**, close to the ENZ region of the TCO, the electric field is confined mostly in the TCO while the electric field out of TCO decreases (blue curve at **Fig. 3**).

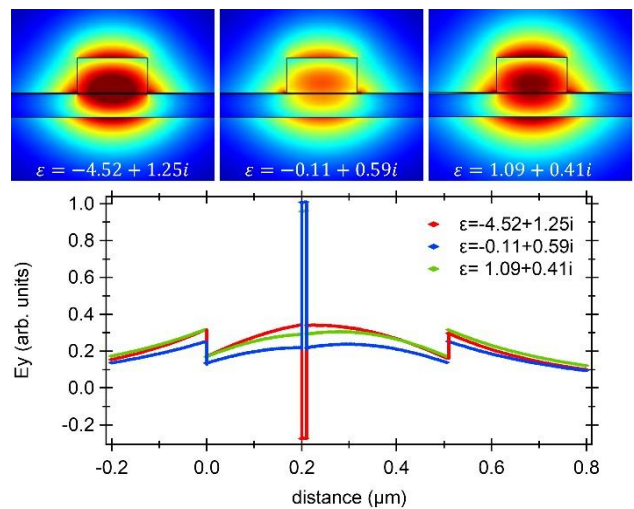

**Figure 3.** Electric field distribution in the SiN waveguide with ITO for different value of ITO permittivity.

In consequence, the mode power attenuation is the highest at ENZ region as observed from **Fig. 4a** for ITO and decreases fast when moving out from ENZ point. The absorption curve reminds well-known bell shape. Depending on the power coupled to the SiN rib waveguide and the carrier concentration of ITO that is part of the waveguide, a device can operate in a bistable region with two different stable levels of transmitted output power for the same input (**Fig. 4b-e**) [15, 16]. Thus, it can serve as a memristor that mimics the biological synaptic response and allow to co-locate both processing and storage. Memristors have opened new doors to integrated circuits as it allows to actively modulate electrical or optical signals and hold memory states comparable to synaptic activity in the brain [18-21].

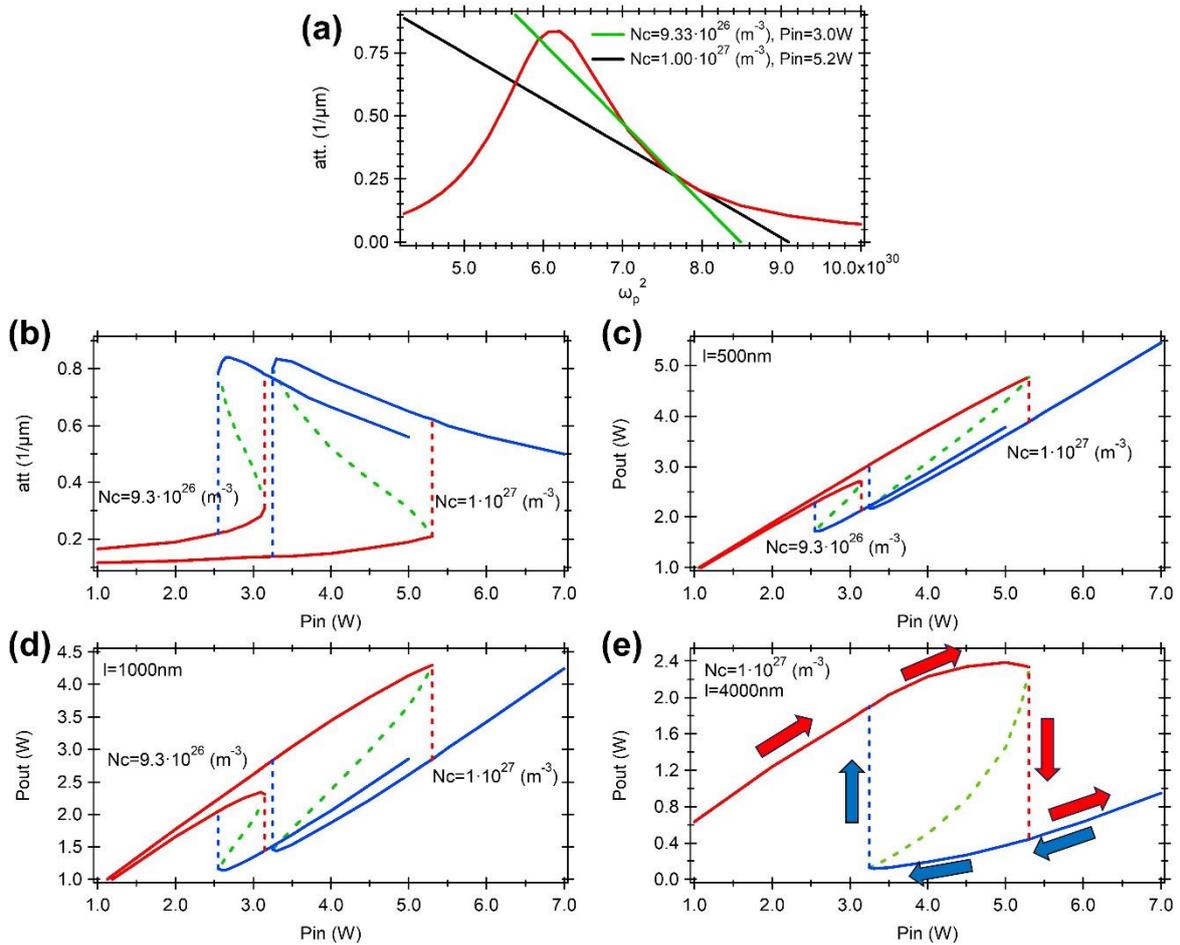

**Figure 4.** (a) Illustration of bistability and switching for different optical powers in the waveguide and under a different carrier concentration of ITO – $N_c$=0.93·10$^{27}$ m$^{-3}$ and $N_c$=1.0·10$^{27}$ m$^{-3}$. The mechanism of bistability was explained in detail in ref. **15, 16**. (b) Absorptive loss as a function of the propagating power exhibiting hysteresis and manifesting all-optical bistability. (c, d, e) Input–output characteristics of the photonic bistable device of (b) 500 nm, (c) 1000 nm and (d) 1000 nm length for different carrier concentration of 0.93·10$^{27}$ m$^{-3}$ and 1.0·10$^{27}$ m$^{-3}$.

As observed from **Fig. 4**, the optical power required to move into a bistable region for SiN photonic waveguide with ITO is pretty high in a range of few Watts. The higher carrier concentration provides wider bistable region but at the cost of input optical power that arises. Simultaneously, the longer device no influence the bistability region range, however it highly influences the output power contrast between low and high transmission levels. And, with longer devices the absorption arises for both transmissions. For higher carrier concentration the bistability region ranges from 3.25 W to 5.3 W while for lower carrier concentration from 2.55 W to 3.15 W. For a device length of *l*=500 nm the output power difference in a low and high transmission level ranges from 3 W to 4.8 W for a high transmission level and from 2.2 W to 3.9 W for a low transmission level. In comparison, for a longer device of *l*=4000 nm it ranges from 1.9 W to 2.35 W and from 0.13 W to 0.46W for high and low transmission levels, respectively. For a shorter device a difference is around 1.8 W while for longer device is around 0.35-0.45 W. As observed, the change of input optical power from 3.25 W to 5.3 W causes only small change in the output power for longer device – both transmission lines flatten out.

The operation conditions of the proposed photonic device can be changed when ITO is replaced by other TCO material (**Fig. 5**). For the same plasma frequency, the power required to operate in a bistable region drops from 3.25-5.30 W for structure with ITO (**Fig. 4**) to only 0.18-0.37 W for structure with In:CdO (**Fig. 6**). It is over 18 times reduction in the input power required to move into a bistability region. Even GZO can be helpful to reduce the power while working into a lower plasma frequency (**Fig. 5**). It means that even with lower carrier concentration in the GZO, the power can be reduced over few times compared to a device with ITO.

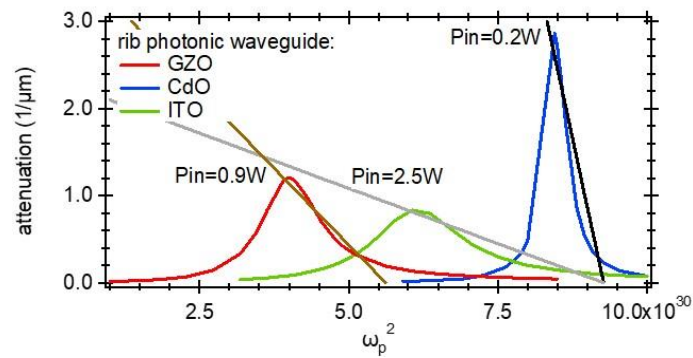

**Figure 5.** Illustration of bistability and switching as power changes in the waveguide for different TCO materials – GZO, ITO, and In:CdO.

Furthermore, the absorption curve of a proposed device should be as narrow and steep as possible to ensure low power consumption (**Fig. 5**). It is directly related to the TCO material properties what was mentioned previously – the steeper a real part of permittivity close to the ENZ region the narrower absorption curve. And lower imaginary part of permittivity translates on higher absorption contrast as absorption highly arises only if the TCO material works close to the ENZ region (**Fig. 2**). From **Fig. 5** and **Fig. 2** can be deduced that ultra-low loss TCO materials with sharp index dispersion in the ENZ range are preferred as they offer sharp and narrow absorption curve of a device what reduces a power required for switching.

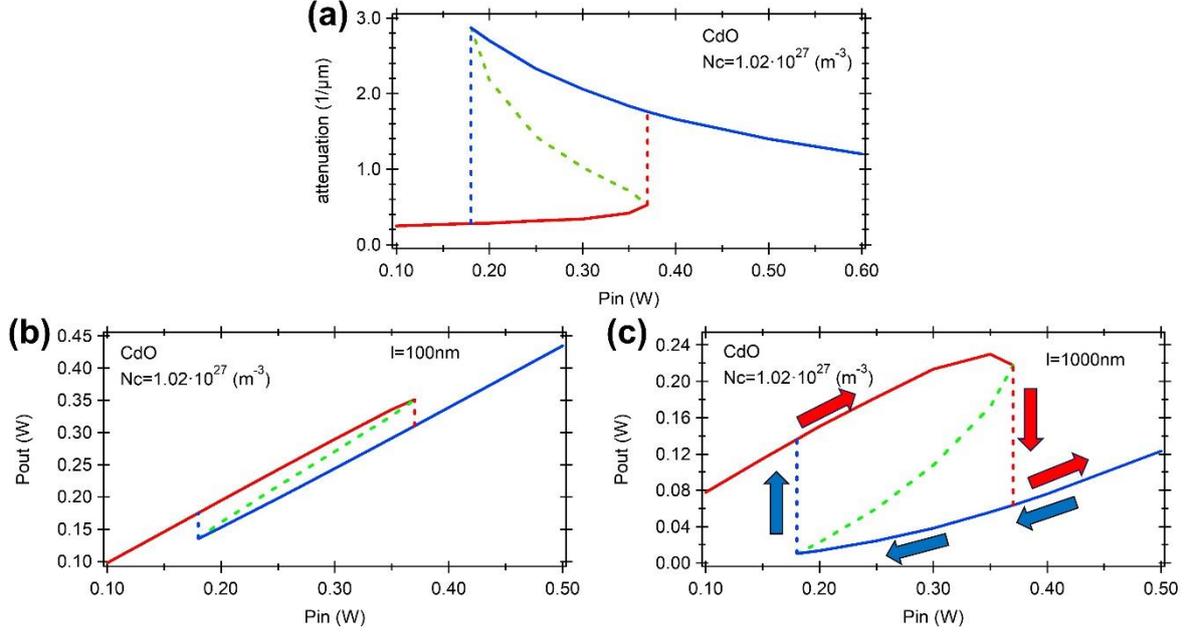

**Figure 6.** (a) Absorptive loss as a function of the propagating power for Indium doped CdO and for carrier concentration $N_c$=1.02·$10^{27}$ $m^{-3}$. (b, c) Input–output characteristics of the photonic bistable device of (b) 100 nm and (c) 1000 nm length for carrier concentration of 1.02·$10^{27}$ $m^{-3}$.

For a device with *l*=1000nm long In:CdO, a difference between low transmission level and high transmission level in a bistable region of a device changes from 10 mW to 135 mW for an input power of 180 mW and from 63 mW to 220 mW for an input power of 370 mW. Higher contrast is possible with longer devices but at the cost of output power that drops for longer devices. Here we operate at telecom wavelength of 1550 nm, thus highly doped In:CdO is required, however, we can reduce a doping level when working close to the ENZ wavelength of 1820 nm (Fig. 2a).

The proposed photonic bistable device can serve as a building block for complex photonic neural networks. The proper choice of TCO materials and operation wavelength allow to define the operation conditions of the device while a design allow to enhance interaction of light with TCO material. To imitate brain performances such devices should be arranged in more complex architectures that can serve for a neuromorphic computing based on a photonic platform.

**Dual-mode operation**

By playing simultaneously with both electrical and optical switching [**40, 41**] or only all-optical switching but under both interband and intraband absorption (two light sources at different wavelength) we can take a full advantage of the switching possibilities of TCO materials. As observed from **Fig. 7**, even when we change simultaneously or step by step a carrier concentration in TCO and effective mass through coupling a light to the TCO, we can still stay at the same value of permittivity. From this point, the performance of a device does no change (points E and F and solid line in **Fig. 7a**). However, when we increase an effective mass of the TCO through coupling a short pulse of light to a device and simultaneously increase a carrier concentration in the TCO through either intraband pump or electric voltage, and when a light pulse is off, a device transfer from a high loss regime, $\varepsilon_r$≈0, to a low loss regime, $\varepsilon_r$≈-1.8 (points A and B and dotted line in **Fig. 7a**). In a contrary, by working in a different parameters range, we can transfer from a low loss regime $\varepsilon_r$≈-2.0 (point C) to a high loss regime $\varepsilon_r$≈0 (point D) by coupling a light to a device and simultaneously either applying a short electrical pulse to the TCO or coupling a short optical pulse to a device (dashed line in **Fig. 7a**).

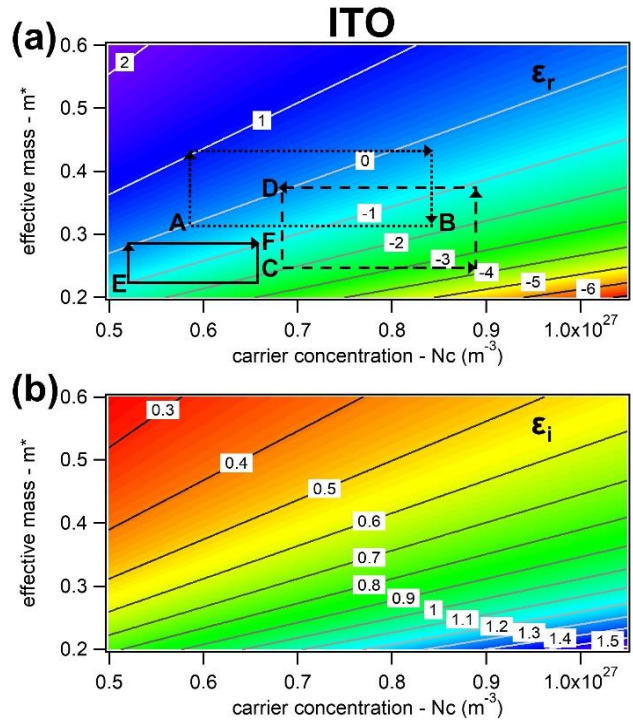

**Figure 7**. (a) Real and (b) imaginary part of permittivity map for different carrier concentration and effective mass.

Thus, transparent conductive oxides (TCOs) open new possibilities in both photonic integrated circuits (PIC) and neuromorphic photonics that can provide a lot of freedom in a design and can bring a network to the next operational level.

**Biological brain**

As the goal of the neuromorphic computing is to mimic a behavior of a biological brain we should, at first, recall how a signal is processed in biological systems [**3, 4**]. In a brain, two types of synaptic integration take place, and both of them are very essential for a signal processing. First, the spatial summation – the process in which synaptic potential generated at many different synapses on a dendrite of the same neuron are added together at the soma region. Second, the temporal summation – the process in which many synaptic potentials generated at the same synapse are added together if they occur in rapid succession. Thus, it requires high-frequency presynaptic activity to summate all the postsynaptic responses. Going into more details – in the absence of any signals in the neuron, the membrane of the individual neuron stays at so-called the resting potential. To generate an action potential the membrane potential must be reduced below threshold what is called depolarization. As the depolarization enhances a cell's ability to generate an action potential, it is excitatory. It has been already mentioned that to achieve the necessary depolarization the synapses must be stimulated at high frequencies. Furthermore, to achieve a significant spatial summation enough synapses must be active simultaneously. This second requirement in a biology is called cooperativity as many coactive synapses must cooperate to produce enough depolarization to cause long-term potentiation, *i.e.*, activity. To achieve a sufficient temporal summation, the individual presynaptic potential must persist long enough to maintain depolarization and even deepen it before the next presynaptic potential arrives. Thus, it defines the membrane time constant that determines the time course of the synaptic potential and thus controls temporal summation. In a human brain, a time constant is in the range of 1-15 ms. In consequence, the neurons with a larger membrane time constant have a greater capacity for temporal summation as there is higher probability that two consecutive signals from presynaptic neuron will summate and bring the membrane to the threshold for an action potential.

**Device performance in neural networks**

For the proposed structure, the optical signal corresponds to a biological equivalent of action potential in neurons while a thermalization time of electrons in TCO corresponds to the membrane time constant. The membrane time constant defines how long the depolarization is maintained by the neurons while a thermalization time of electrons defines a time needed for excited electrons to return to its initial unexcited state. Consequently, while a depolarization defines a membrane potential that is reduced below threshold, its equivalent in proposed device defines the lower output optical power for a given input optical power for an all-optical switching mechanism or higher output optical power for the same carrier concentration for an electrical switching mechanism. Similarly to the biological counterpart, a temporal summation in the proposed device that is based on the TCO materials require high-frequency input optical pulses to summate all the signals. A time constant between consequent optical pulses should be shorter than a thermalization time of electrons in the TCO so the energy provided to the electrons from the next optical pulse should give rise to further increases of the energy of the electron gas. For a time constant between consequent optical pulses longer than an electron-lattice relaxation time of electrons into TCO, the electrons excited by a first optical pulse return to its initial unexcited state before the next optical pulse arrives.

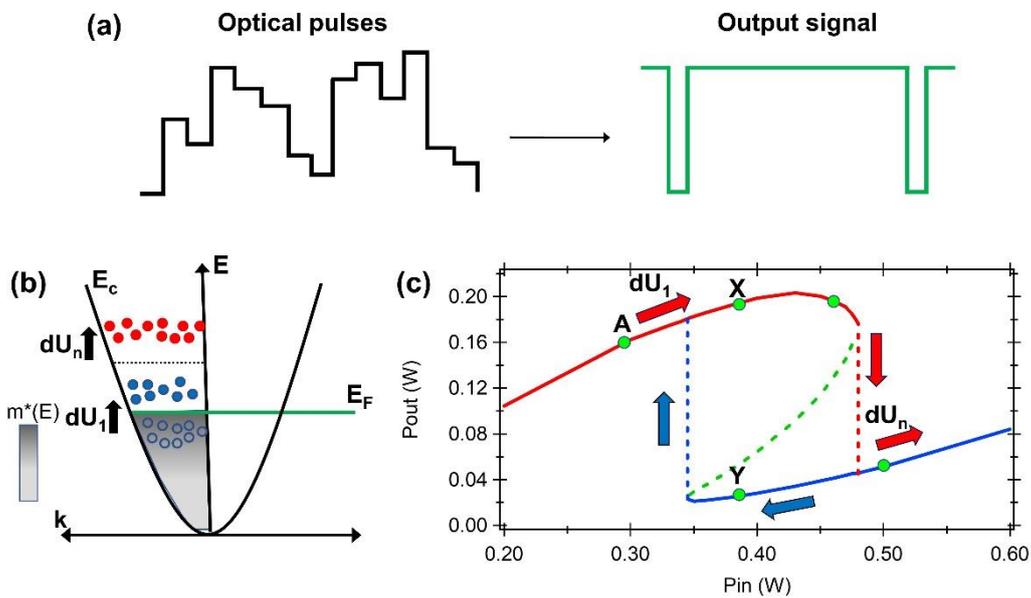

**Figure 8.** (a) An optical pulse train from waveguide before and after a proposed device. (b) Absorbed pump energy $U_1$, ..., $U_n$ under consequent pulses increases electron energy and, thus, the electrons effective mass $m^*(E)$ and (c) operation principles.

When the consequent pulses are high enough and are combined in the integration area in a time shorter than a thermalization time of electrons in TCO, each pulse slightly heats up electrons and move it higher in the conduction band. The output power follows the red curve as shown in **Fig. 8**. However, when the combined optical power exceeds the threshold, the optical transmission drops and now follows the blue curve. When the optical pulses delivered to the device decrease or if a distance between consequent pulses exceeds a thermalization time of electrons in TCO, the electrons can thermalize and return to its initial energy level, thus in consequence, the effective mass decreases and transmission drops to lower level for the same input power (points X and Y in **Fig. 8**). Further decreases of the optical input power and thus the electron temperature reset a device and moves its back to its initial state indicated by point A. In this arrangement, the integration of pulses can be both in spatial and temporal domains where the pulses from other neurons can be combined into a single waveguide using wavelength division multiplexing (WDM). In this case, as the switching of the TCO occurs only above a certain threshold value, the neuron only stays at low output power if the weighted sum of the input optical power exceeds this threshold. Thus, the system naturally emulates the basic integrate-and-fire functionality of a biological neuron but in inverse schema – system stays in low power only when the threshold is reached. This artificial neuron can integrate over the optical power and over time what make it very similar to a biological neuron.

**Conclusion**

For a first time we have examined a bistable device on the low-loss nitride-rich silicon platform with the TCO active materials arranged in the photonic rib waveguide for application in artificial neural networks. Different TCO materials were examined showing that significant reduction in optical power can be achieved under proper choice of material. The proposed photonic device can serve as both a linear weight for a single photonic signal and a simultaneously spatial and temporal summation unit integrating many photonic signals. Furthermore, depending on the overall summated signal value the proposed device can keep history about previous state and thus can serve as a memristor what bring it closer to the brain. The proposed device can be easily integrated with the photonic SiN waveguides serving as an interconnector with a coupling efficiency exceeding even 95 %. Furthermore, both materials *i.e.*, silicon nitride and transparent conductive oxides are CMOS-compatible and are characterized by very low losses what open new possibilities for a further development of neural networks.

**Acknowledgements**

J.G. thanks the "ENSEMBLE3 - Centre of Excellence for Nanophotonics, advanced materials and novel crystal growth-based technologies" project (GA No. MAB/2020/14) carried out within the International Research Agendas program of the Foundation for Polish Science co-financed by the European Union under the European Regional Development Fund and the European Union's Horizon 2020 research and innovation program Teaming for Excellence (Grant Agreement No. 857543) for support of this work.